\documentclass[%
 aip,
cp,  
 amsmath,amssymb,
 reprint,%
nofootinbib
]{revtex4-2}

\usepackage{graphicx}
\usepackage{dcolumn}
\usepackage{bm}
\usepackage{color}
\usepackage{combelow}

\usepackage[utf8]{inputenc}
\usepackage[T1]{fontenc}
\usepackage{mathptmx} 
\usepackage{ulem}
\usepackage[colorlinks=true,linkcolor=blue,citecolor=blue, urlcolor=blue]{hyperref} 

\def\xT{{\mathbf{x}_\perp}}
\def\wt{{\tilde{w}}}

\begin{document}

\title{Attractors for Flow Observables in $2+1$D Bjorken Flow}

\author{Victor E. Ambru\cb{s}} 
 \email[Corresponding author: ]{victor.ambrus@e-uvt.ro}

\affiliation{Department of Physics, West University of Timi\cb{s}oara, \\
Bd.~Vasile P\^arvan 4, Timi\cb{s}oara 300223, Romania}
\author{S\"oren Schlichting}%
\affiliation{Fakultät für Physik, Universität Bielefeld, D-33615 Bielefeld, Germany}

\author{Clemens Werthmann}
\affiliation{Fakultät für Physik, Universität Bielefeld, D-33615 Bielefeld, Germany}

\date{\today} 

\begin{abstract}
We examine the capabilities of second-order Israel-Stewart-type hydrodynamics to capture the
early-time behaviour of the quark-gluon plasma created in heavy-ion collisions. We point out
that at very early times, the dynamics of the fireball is governed by the local 0+1-D Bjorken flow
attractor due to the rapid expansion along the longitudinal direction. Discrepancies between
hydrodynamics and kinetic theory in this far-from-equilibrium regime leads to disagreement at the
level of late-time observables, such as elliptic flow. We show that rescaling the initial energy-density
profile for hydrodynamics accounts for such discrepancies, restoring agreement with kinetic theory
for large opacities (small shear viscosity / large system size / high energy).
\end{abstract}

\maketitle

\section{Introduction}

The transition of strongly-interacting matter from the ordinary hadronic phase to a deconfined phase dubbed the Quark Gluon Plasma (QGP) has been theoretically predicted in the early '70s based on the asymptotic freedom property of the strong interaction at high energies. Experimental evidence for the existence of the QGP phase came from heavy-ion collision experiments that reported observables based on collective behaviour, such as elliptic flow, typically expected to emerge in a strongly-correlated fluid. While initially extremely hot, the QGP formed shortly after the cools down due to longitudinal and transverse expansion, eventually transitioning to the confined, hadronic phase when its temperature drops below the critical temperature, estimated by lattice QCD studies at $k_B T_c \simeq 150\ {\rm MeV}$.

In its initial stages, the constituents of the QGP are in a far-from-equilibrium state. In the local rest frame of the plasma, the created gluons have mostly transverse momenta and the overall fluid exhibits nearly-vanishing longitudinal pressure, $P_L \simeq 0$. At initial times, the longitudinal expansion dominates the system's dynamics, and the ratio $P_L / P_T$ between the longitudinal and transverse pressures is close to zero. After this effectively free-streaming stage, the longitudinal expansion timescale, proportional to the Bjorken time $\tau$, becomes comparable to the interactions timescale $\tau_R$ and the system tends to equilibrate as $P_L / P_T$ increases towards unity. In small systems or at low energies, equilibration is interrupted by transverse expansion and the fluid remains out of equilibrium during its entire lifetime. 

As pointed out in the seminal work of Heller and Spa{\l}inski \cite{Heller:2015dha}, thermal equilibration is preceded by the phenomenon of hydrodynamization, which is related to the decay of so-called non-hydrodynamic modes onto a universal attractor solution. Hydrodynamization also marks the end of the free-streaming phase, such that after this point, the dynamics of the fluid can be modeled accurately using relativistic hydrodynamics. In our work, we study the dynamics of the system during the pre-equilibrium stage using both kinetic theory and relativistic hydrodynamics and assess 
the applicability of the latter for the early-stage evolution. More details can be found in Ref.~\cite{Ambrus:2022}.


\section{Initial state and observables}

We work under the assumption of boost invariance along the longitudinal ($z$) direction, which is expected to hold in the mid-rapidity region of heavy-ion collisions \cite{Bjorken:1982qr}. We employ the Bjorken coordinates $x^\mu = (\tau, \xT, \eta_s)$, where $\tau = \sqrt{t^2 - z^2}$ and $\eta_s = \tanh^{-1}(z/t)$. The fluid four-velocity that preserves boost invariance reads $u^\mu = (u^\tau, \mathbf{u}_\perp, 0)$.

At initial time, we consider that the system is described by a diagonal energy-momentum tensor, $T^{\mu\nu}(\tau_0, \xT) = {\rm diag}(\epsilon_0, P_T, P_T, \tau^{-2} P_L)$, where the longitudinal and transverse pressures are initialized according to the attractor solution
discussed in the next section.
The initial energy density $\epsilon_0 = \frac{1}{\tau_0} dE_\perp^0 / d\eta d^2\mathbf{x}_\perp$ is computed based on the transverse-plane energy density per unit rapidity, $dE_\perp^0 / d\eta d^2\mathbf{x}_\perp$, representing a
realistic average initial condition for the 30-40\% most central Pb-Pb collisions at $\sqrt{s_{NN}} = 5.02\ {\rm TeV}$, typical of LHC experiments (see Ref.~\cite{Borghini:2022iym} for the event generator details).
The system is characterized by the total energy per unit rapidity $dE^0_\perp / d\eta$, effective radius $R$ and eccentricities $\epsilon_n$, computed via 
\begin{align}
 \frac{dE^0_\perp}{d\eta} &= \int_{\mathbf{x}_\perp} \tau_0 \epsilon_0, & 
 R^2 \frac{d E^{0}_\perp}{d\eta} &= \int_{\mathbf{x}_\perp} \tau_0 \epsilon_0 
 \mathbf{x}_\perp^2, &
 \epsilon_n &= -\frac{\int_{\mathbf{x}_\perp} \tau_0 \epsilon_0 x_\perp^n \cos\left[n(\phi_x-\Psi_n)\right]}
 {\int_{\mathbf{x}_\perp} \tau_0 \epsilon_0 x_\perp^n},
\end{align}
where $\int_{\mathbf{x}_\perp} \equiv \int d^2\mathbf{x}_\perp$ denotes the integration over the transverse plane. For our particular choice of profile, $d E^{0}_\perp / d\eta = 1280\ {\rm GeV}$, $R = 2.78\ {\rm fm}$, $\epsilon_2 = 0.42$, $\epsilon_4 = 0.21$ and $\epsilon_6 = 0.09$, while the odd-order eccentricities vanish.

The time evolution of the initial conditions discussed above is modeled using two approaches, namely kinetic theory and relativistic hydrodynamics.
In the kinetic theory description, we employ the relativistic Boltzmann equation in the relaxation-time approximation (RTA) of Anderson and Witting \cite{Anderson:1974}, 
\begin{align}
 p^\mu \partial_\mu f &= -\frac{p_\mu u^\mu}{\tau_R} (f - f_{eq}), & 
 f_{eq} &= \frac{1}{\exp(p \cdot u / T) - 1}, &
 \tau_R &= \frac{5 \eta}{s T},
\label{eq:RTA}
\end{align}
where $f$ is the averaged on-shell phase-space distribution, $p^\mu = (E_{\mathbf{p}}, \mathbf{p})$ is the four-momentum for massless particles with energy $E_{\mathbf{p}} = |\mathbf{p}|$ and $\tau_R$ is the relaxation time. The Bose-Einstein equilibrium distribution $f_{eq}$ is characterized by the temperature $T = (\epsilon / a)^{1/4}$, where $a = \frac{\pi^2}{30} \nu_{\rm eff}$ and $\nu_{\rm eff} = 42.25$ is the effective number of degrees of freedom for a free gas of gluons and $2+1/2$ massless quark flavors. The energy density $\epsilon$ and four-velocity $u^\mu$ are obtained via Landau matching, 
$T^{\mu\nu} u_\nu = T^{\mu\nu}_{eq} u_\nu = \epsilon u^\mu$.
We consider a constant shear viscosity to entropy density ratio $\eta / s$, with $s = (\epsilon + P) / T$.
Equation~\eqref{eq:RTA} is solved using the relativistic lattice Boltzmann method, as described in Ref.~\cite{Ambrus:2022}. 


The hydrodynamic description starts from the following decomposition of the energy-momentum tensor
\begin{equation}
 T^{\mu\nu} = \epsilon u^\mu u^\nu - P \Delta^{\mu\nu} + \pi^{\mu\nu},
\end{equation}
where $\epsilon = 3P$, $g^{\mu\nu}$ is the inverse metric tensor, $\Delta^{\mu\nu} = g^{\mu\nu} - u^\mu u^\nu$ is the projector on the hypersurface orthogonal to $u^\mu$, while $\pi^{\mu\nu}$ is the shear-stress tensor. Imposing the conservation of $T^{\mu\nu}$ leads to evolution equations for $\epsilon$ and $u^\mu$,
\begin{align}
 \dot{\epsilon} + (\epsilon + P) \theta - \pi^{\mu\nu} \sigma_{\mu\nu} &= 0, &
 (\epsilon + P) \dot{u}^\mu - \nabla^\mu P
 + \Delta^\mu{}_\lambda \partial_\nu \pi^{\lambda\nu} &= 0,
 \label{eq:hydro_cons}
\end{align}
where $\dot{\epsilon} = u^\mu \partial_\mu \epsilon$, $\theta = \partial_\mu u^\mu$, $\sigma_{\mu\nu} = \nabla_{\langle \mu} u_{\nu \rangle}$, $\nabla_\mu = \partial_\mu - u_\mu u^\nu \partial_\nu$
and we understand that $\partial_\mu$ acts as a covariant derivative with respect to the curvilinear Bjorken coordinates. Furthermore, $A^{\langle \mu\nu \rangle} = \Delta^{\mu\nu}_{\alpha\beta} A^{\alpha\beta}$ and $\Delta^{\mu\nu}_{\alpha\beta} =
\frac{1}{2} (\Delta^\mu_\alpha \Delta^\nu_\beta + \Delta^\nu_\alpha \Delta^\mu_\beta) - \frac{1}{3} \Delta^{\mu\nu} \Delta_{\alpha\beta}$.
In ideal hydrodynamics, $\pi^{\mu\nu} = 0$ and the evolution equations \eqref{eq:hydro_cons} are closed. In the M\"uller-Israel-Stewart (MIS) theory of viscous hydrodynamics, the shear-stress tensor $\pi^{\mu\nu}$ evolves according to
\begin{equation}
 \tau_\pi \dot{\pi}^{\langle \mu\nu \rangle} + \pi^{\mu\nu} = 2\eta \sigma^{\mu\nu} + 
 2\tau_\pi \pi^{\langle\mu}_\lambda \omega^{\nu\rangle \lambda} 
 - \delta_{\pi\pi} \pi^{\mu\nu} \theta - 
 \tau_{\pi\pi} \pi^{\lambda\langle \mu}\sigma^{\nu\rangle}_\lambda + \phi_7 \pi_\alpha^{\langle\mu}\pi^{\nu\rangle\alpha}.
 \label{eq:hydro_pi}
\end{equation}
The transport coefficients appearing above are selected to ensure compatibility with RTA, namely
\begin{align}
 \eta &= \frac{4}{5} \tau_\pi P, &
 \tau_\pi &= \tau_R, &
 \delta_{\pi\pi} &= \frac{4\tau_\pi}{3}, &
 \tau_{\pi\pi} &= \frac{10\tau_\pi}{7}, &
 \phi_7 &= 0.
 \label{eq:hydro_tcoeffs}
\end{align}
The numerical solution of Eqs.~\eqref{eq:hydro_cons} and \eqref{eq:hydro_pi} is obtained using the vHLLE code \cite{Karpenko:2013wva}, modified as explained in Ref.~\cite{Ambrus:2022}.

The observables of our study are the transverse-plane energy per rapidity $dE_{\rm tr} / d\eta$ and elliptic flow coefficient $\varepsilon_p$,
\begin{align}
 \frac{d E_{\rm tr}}{d\eta} &= \tau \int_{\xT} (T^{xx}+T^{yy}), &
 \varepsilon_p e^{2i \Psi_p} &= \frac{\int_{\xT} (T^{xx}-T^{yy}+2iT^{xy})}{\int_{\xT} (T^{xx}+T^{yy})},
 \label{eq:obs} 
\end{align}
where $\Psi_p$ is an event-plane angle.

\section{Bjorken attractor}

During its evolution, the system encounters three different timescales. At early times, the dynamics of the system is dominated by the longitudinal expansion, whose time scale is related to the inverse of the expansion scalar $\theta^{-1} \simeq \tau$, thus becoming arbitrarily short as $\tau \rightarrow 0$. A second time scale is set by the relaxation time $\tau_R$ and is related to the equilibration of the fluid. Finally, the system size $R$ sets the third time scale, related to the onset of transverse expansion. In this section we focus on the regime when $\tau / R \ll 1$, when transverse expansion plays a subleading role and as a first approximation, it can be neglected. 
Under such circumstances, 
the time evolution is described by the $0+1$-D Bjorken model, by which the four-velocity reduces to $u^\mu = (1,0,0,0)$. The energy-momentum tensor remains diagonal, $T^\mu_\nu = {\rm diag}(\epsilon, -P_T, -P_T, -P_L)$, while the hydrodynamical equations \eqref{eq:hydro_cons} and \eqref{eq:hydro_pi} reduce to
\begin{align}
 \tau \frac{d \epsilon}{d\tau} + \frac{4}{3} \epsilon + \pi_d &= 0, &
 \tau \frac{\partial \pi_d}{\partial \tau} + \left(\lambda + \frac{4\pi \tilde{w}}{5} + \frac{2 \pi \tilde{w}}{5} \phi_7 \pi_d \right) \pi_d + \frac{16 \epsilon}{45} &= 0, & 
 \lambda &= \frac{1}{\tau_\pi} (\delta_{\pi\pi} + \tau_{\pi\pi} / 3).
 \label{eq:bjorken_hydro}
\end{align}
In the above, $\pi_d = \tau^2 \pi^{\eta\eta} = \frac{2}{3}(P_L - P_T)$, or equivalently, $P_T = P - \pi_d / 2$ and $P_L = P + \pi_d$, while $\eta = \frac{4}{5} \tau_\pi P$, $\lambda = 38/21$ and $\phi_7 = 0$ [see Eq.~\eqref{eq:hydro_tcoeffs}]. The definition of the conformal parameter $\tilde{w}$ and its evolution equation are
\begin{align}
 \tilde{w} &= \frac{5\tau}{4\pi \tau_\pi} = \frac{\tau T}{4\pi \eta / s}, &
 \tau \frac{d\wt}{d\tau} &= \left(\frac{2}{3} - \frac{f_\pi}{4}\right) \wt.
 \label{eq:wt}
\end{align}

\begin{figure}
\begin{center}
\begin{tabular}{cc}
 \includegraphics[width=0.48\linewidth]{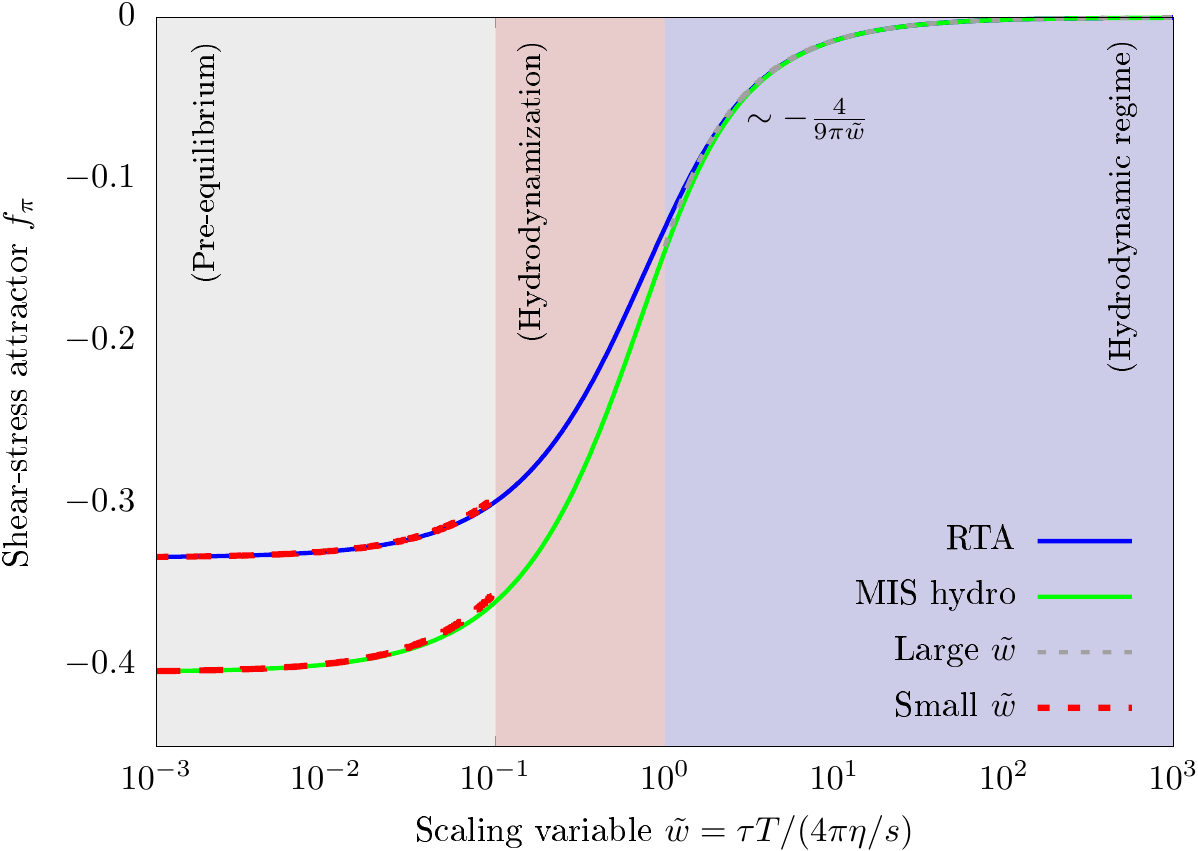} &
 \includegraphics[width=0.48\linewidth]{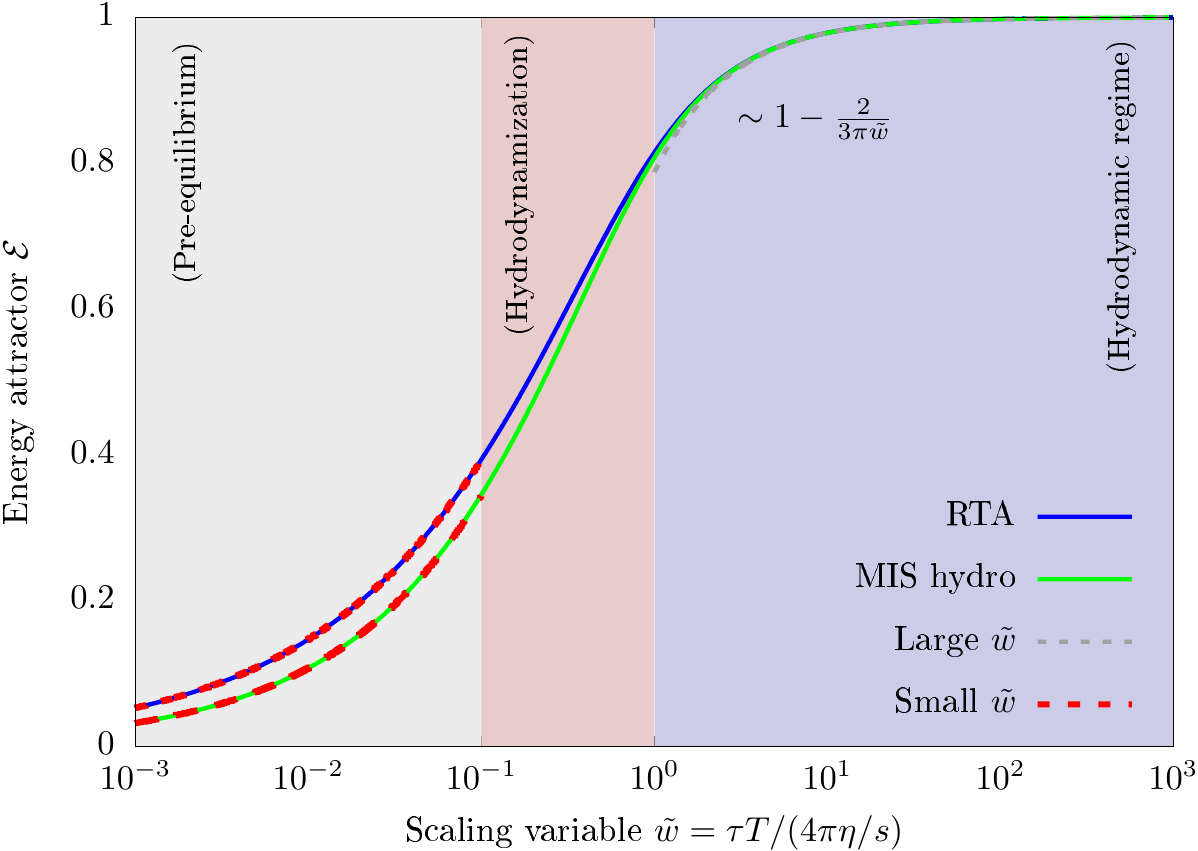}
\end{tabular}
\caption{
Attractor functions $\mathcal{E}$ and $f_\pi$ for the energy density and shear-stress coefficient, obtained numerically for RTA (blue lines) and hydrodynamics (green lines). The asymptotic approximations \eqref{eq:bjorken_early_fpi} and \eqref{eq:bjorken_early_E} for small $\tilde{w}$ are shown with dotted red lines. The large-$\tilde{w}$ approximations in Eq.~\eqref{eq:bjorken_late} are shown with dotted gray lines.
\label{fig:attractor}
}
\end{center}
\end{figure}

As is well known \cite{Heller:2015dha}, the function $f_\pi = \pi_d / \epsilon$ admits an attractor solution which depends only on $\tilde{w}$.
A particular feature of the attractor solution $f_\pi(\tilde{w})$ is that it remains regular at early times ($\wt \ll 1$), 
\begin{equation}
 f_\pi(\tilde{w} \ll 1) = f_{\pi;0} + f_{\pi; 1} \tilde{w} + O(\tilde{w}^2),
 \label{eq:bjorken_early_fpi}
\end{equation}
where the values $f_{\pi;0}$ and $f_{\pi;1}$ depend on the theory under consideration. In RTA, we have $f_{\pi;0} = -1/3$ and $f_{\pi;1} \simeq  0.370$, while for MIS hydrodynamics, $f_{\pi;0} \simeq -0.404$ and $f_{\pi;1} \simeq 0.495$ \cite{Ambrus:2022}.

Replacing $\pi_d = f_\pi \epsilon$ and $\tau^{4/3} \epsilon = \tau_0^{4/3} \epsilon_0 \mathcal{E}(\tilde{w}) /  \mathcal{E}(\tilde{w}_0)$, Eqs.~\eqref{eq:bjorken_hydro} and \eqref{eq:wt} show that the energy attractor $\mathcal{E}(\tilde{w})$ satisfies
\begin{equation}
 \tilde{w} \left(\frac{2}{3} - \frac{f_\pi}{4}\right) \frac{d \mathcal{E}}{d \tilde{w}} + f_\pi \mathcal{E} = 0.
 \label{eq:bjorken_Eprime}
\end{equation}
At early times, when Eq.~\eqref{eq:bjorken_early_fpi} is valid, $\mathcal{E}$ has the following behavior:
\begin{align}
 \mathcal{E}(\tilde{w} \ll 1) &= C_{\infty}^{-1} \tilde{w}^{\gamma} [1 + \mathcal{E}_1 \tilde{w} + O(\tilde{w}^2)], &
 \gamma &= \frac{4 f_{\pi;0}}{f_{\pi;0} - \frac{8}{3}}, &
 \mathcal{E}_1 &= -\frac{\frac{32}{3} f_{\pi;1}}{(f_{\pi;0} -\frac{8}{3})^2},
 \label{eq:bjorken_early_E}
\end{align}
where $C_\infty$ is an integration constant (discussed further down). In RTA, $\gamma = 4/9$ and $\mathcal{E}_1 \simeq -0.439$, while in the MIS theory we have $\gamma = (\sqrt{505}-13) / 18 \simeq 0.526$ and $\mathcal{E}_1 \simeq -0.699$.

At late times, when $\tau \gg \tau_R$, the fluid has undergone equilibration to a significant degree and both RTA and MIS hydrodynamics are described by Eqs.~\eqref{eq:bjorken_hydro}.
In this case, the large $\tilde{w}$ limits of $f_\pi$ and $\mathcal{E}$ read for both theories as follows:
\begin{align}
 f_\pi(\tilde{w} \gg 1) &= -\frac{4}{9\pi \tilde{w}} + O(\tilde{w}^{-2}),&
 \mathcal{E}(\tilde{w} \gg 1) &= 1 - \frac{2}{3\pi \tilde{w}} + O(\tilde{w}^{-2}).
 \label{eq:bjorken_late}
\end{align}
While no integration constant can be specified for $f_\pi$, the late-time asymptotics $\lim_{\tilde{w} \to \infty} \mathcal{E}(\tilde{w}) = 1$ fixes the constant $C_\infty$ appearing in Eq.~\eqref{eq:bjorken_early_E} to $C_\infty \simeq 0.88$ for RTA and $C_\infty \simeq 0.80$ for MIS hydrodynamics \cite{Giacalone:2019ldn}.

The attractor functions $\mathcal{E}(\wt)$ and $f_\pi(\wt)$ are shown in the left and right panels of Fig.~\ref{fig:attractor}, respectively. The solid lines correspond to numerical solutions obtained as described in Ref.~\cite{Ambrus:2022adp}. The asymptotic expansions in Eqs.~\eqref{eq:bjorken_early_fpi} and \eqref{eq:bjorken_early_E} show that the early-time ($\tilde{w} \ll 1$) properties of the RTA and MIS hydrodynamics attractors are different, as confirmed by the dotted red lines. On the other hand, Eq.~\eqref{eq:bjorken_late} indicates that the attractor curves for RTA and MIS hydrodynamics merge as $\tilde{w} \gtrsim 1$, as shown by the dotted gray lines. We thus interpret this late-time regime as the domain of applicability of MIS hydrodynamics. 

\section{Pre-equilibrium evolution and scaling solution}

\begin{figure}
\begin{center}
\begin{tabular}{cc}
 \includegraphics[width=0.48\linewidth]{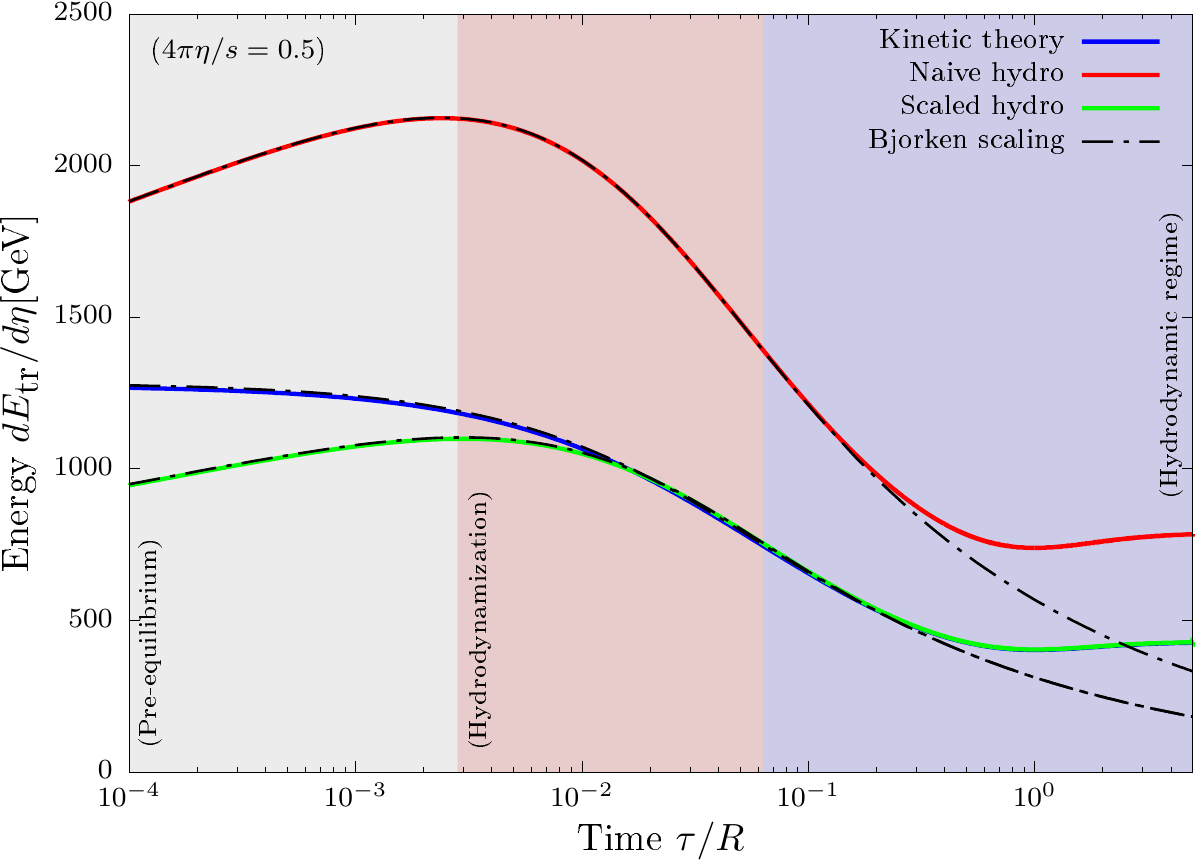} &
 \includegraphics[width=0.48\linewidth]{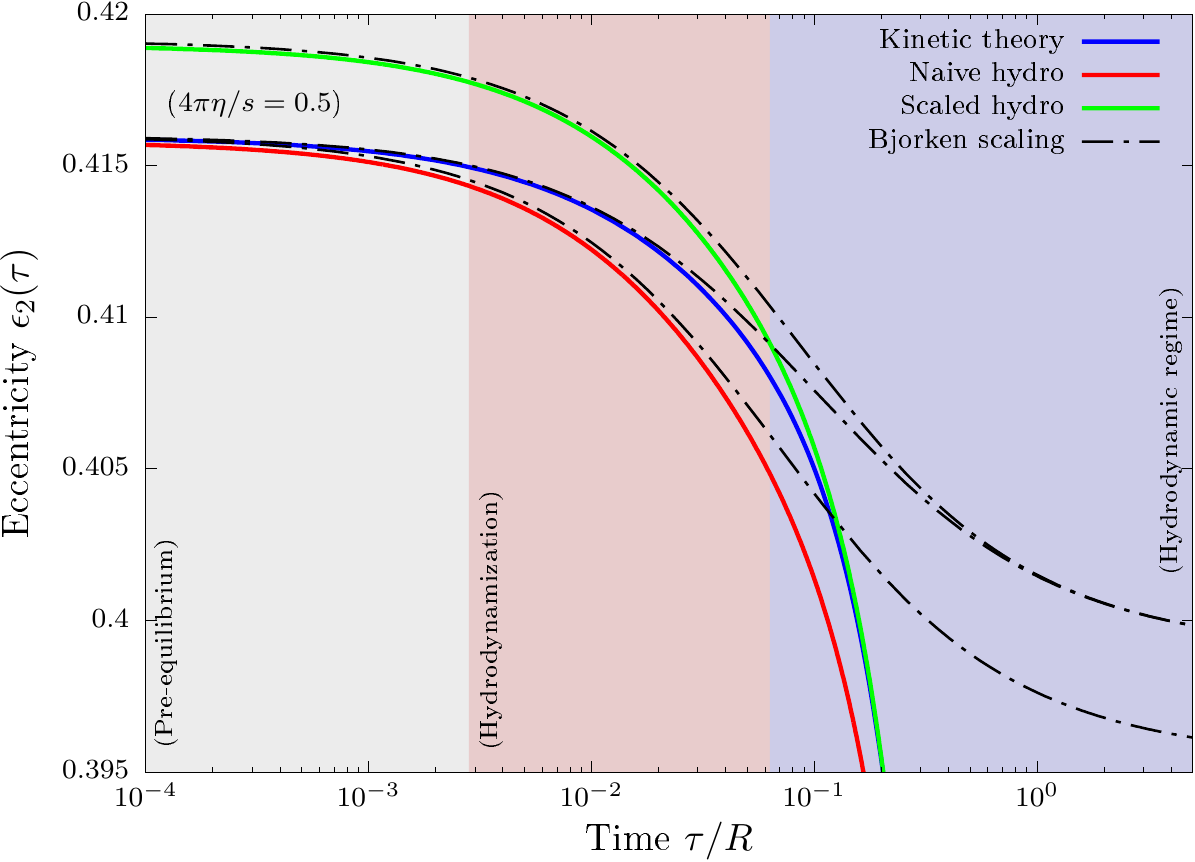}
\end{tabular}
\caption{
Time evolution of the transverse plane energy (left) and elliptical eccentricipty $\epsilon_2$ for a system with $4\pi \eta / s = 0.5$, initialized at $\tau_0 = 10^{-6} R$. 
\label{fig:evolution}
}
\end{center}
\end{figure}

We now investigate the effects of the early-time dynamics on the observables introduced in Eq.~\eqref{eq:obs}. Below the timescale associated to transverse expansion ($\tau \ll R$), transverse dynamics play a subleading role and can thus be ignored. In essence, we will consider that the evolution at each point in the transverse plane is governed by the $0+1$-D Bjorken dynamics. Due to initial inhomogeneities in the transverse energy density profile, the local value of the conformal variable $\wt$ will become position-dependent. For this reason, hotter regions will equilibrate faster than colder regions. This inhomogeneous cooling affects the transverse plane inhomogeneities, in particular the eccentricity coefficients $\epsilon_n$ of the profile. 

At early times, Eq.~\eqref{eq:bjorken_early_E} shows that $\wt \sim \tau^{\frac{2}{3} / (1 - \gamma/4)}$ and $\epsilon(\wt \ll 1) \simeq (\tau_0 / \tau)^{(\frac{4}{3} - \gamma) / (1 - \frac{\gamma}{4})} \epsilon_0$. Substituting this behavior in Eq.~\eqref{eq:obs} gives 
\begin{align}
 \left.\frac{dE_{\rm tr}(\tau)}{d\eta}\right\rfloor_{\text{early time}} &\simeq \left(\frac{\tau_0}{\tau} \right)^{\frac{1}{3}(1 - \frac{9\gamma}{4}) / (1 - \frac{\gamma}{4})} \frac{d E^0_{\rm tr}}{d\eta}, &
 \left.\epsilon_n(\tau)\right\rfloor_{\text{early time}} &\simeq \epsilon_n(\tau_0),
\end{align}
where we used $T^{xx} + T^{yy} = 2P_T = (\frac{2}{3} - f_\pi) \epsilon$.
Evaluating the exponent for RTA, we see that $dE_{\rm tr} / d\eta$ stays constant during the early-time evolution. Conversely, for MIS hydrodynamics, $dE_{\rm tr} / d\eta \propto \tau^{0.07}$ and thus the system undergoes an unphysical 
early-time increase of $dE_{\rm tr} / d\eta$.
In contrast, the eccentricities $\epsilon_n$ are preserved during the early-time evolution for both theories.

Moving now to the late-time dynamics of local Bjorken flow, i.e. $\tilde{w}\gg 1$ while still $\tau \ll R$, we have $\tau^{4/3} \epsilon(\wt \gg 1) \simeq \tau_0^{4/3} \epsilon_0 / \mathcal{E}(\wt_0)$ since $\mathcal{E}(\wt \gg 1) \simeq 1$. Considering that $\tau_0$ is such that $\wt_0 \ll 1$ throughout the system, Eq.~\eqref{eq:bjorken_early_E} can be used to show that 
\begin{equation}
 \epsilon(\wt \gg 1) \simeq \frac{C_\infty}{\tau^{4/3}} \left(\frac{4\pi \eta}{s} a^{1/4}\right)^\gamma
 \left(\tau_0^{(\frac{4}{3} - \gamma) / (1 - \gamma/4)} \epsilon_0\right)^{1 - \gamma/4}.
\end{equation}
Substituting the above into Eq.~\eqref{eq:obs} and considering that, at late times, $T^{xx} + T^{yy} \simeq \frac{2}{3} \epsilon$, we get 
\begin{align}
 \left.\frac{d E_{\rm tr}(\tau)}{d\eta}\right\rfloor_{\text{late time}} &\simeq 
 \frac{2\tau^{-1/3}}{3 C_\infty^{-1}} \left(\frac{4\pi \eta}{s} a^{1/4} \right)^\gamma 
 \tau_0^{\frac{4}{3} - \gamma}
 \int_{\xT} \epsilon_0^{1-\gamma/4},  &
 \left.\epsilon_n(\tau)\right\rfloor_{\text{late time}} &\simeq -\frac{\displaystyle \int_{\xT} x_\perp^n \epsilon_0^{1-\gamma/4} \cos[n(\phi_x - \Psi_n)]}
 {\displaystyle \int_{\xT} x_\perp^n \epsilon_0^{1-\gamma/4}}.
\end{align}
The late-time limit for both observables becomes theory-dependent due to the presence of $\gamma$. 

As mentioned in the introduction, the result obtained under the MIS hydrodynamics evolution cannot be trusted during preequilibrium. For this reason, we aim to counteract the spurious effects of hydrodynamics on the late-time values of our observables by rescaling the initial energy-density profile, such that at late times, $\lim_{\tau \rightarrow \infty} \tau^{4/3} \epsilon_{\rm Hydro} = \lim_{\tau \rightarrow \infty} \tau^{4/3} \epsilon_{\rm RTA}$. This can be achieved by setting the initial energy density in hydrodynamics to
\begin{equation}
 \epsilon_{0,\gamma} = \left[ \left(\frac{4\pi\eta/s}{\tau_0} a^{1/4} 
 \right)^{\frac{1}{2} - \frac{9\gamma}{8}} 
 \left(\frac{C_\infty^{\rm RTA}}{C_\infty^\gamma}\right)^{9/8}
 \epsilon_{0,{\rm RTA}}\right]^{ \frac{8/9}{1 - \gamma/4}}.
 \label{eq:scaling}
\end{equation}

The above analysis is validated in Fig.~\ref{fig:evolution}, where we compare the time evolution of the transverse plane energy $dE_{\rm tr} / d\eta$ (left panel) and ellipticity $\epsilon_2$ (right panel) obtained using kinetic theory (blue lines), naive hydrodynamics initialized with the same initial conditions as RTA (red lines) and scaled hydrodynamics (green lines), initialized according to Eq.~\eqref{eq:scaling}. In all cases, we set the initial conditions at $\tau_0 = 10^{-6} R$ and performed full simulations of a system with $4\pi \eta / s = 0.5$ using the lattice Boltzmann method for RTA \cite{Ambrus:2022,Ambrus:2022adp} and vHLLE for hydrodynamics \cite{Karpenko:2013wva}. On top of the simulation results, dash-dotted black lines show results obtained in the hypothetical scenario where transverse dynamics is completely ignored and the system evolves as an ensemble of independent, point-like $0+1$-D Bjorken flow systems. Due to the unphysical initial rise in $dE_{\rm tr} / d\eta$, the naive hydrodynamics simulation gives a significantly higher energy at late times than RTA. Conversely, the scaled hydrodynamics simulation starts with a significantly lower energy, converging to the RTA curve as the system equilibrates (during hydrodynamization). In the case of the ellipticity, we see that in both RTA and hydrodynamics, the initial value of $\epsilon_2$ is decreased due to inhomogeneous cooling, more in the latter than in the former. For this reason, the late-time asymptotic value of $\epsilon_2$ is lower in naive hydrodynamics than in RTA. Conversely, the initial ellipticity of the scaled hydrodynamics simulation is higher than that of RTA, converging towards the RTA one after equilibration. The discrepancy between the full simulation and the scaling model can be observed for $\tau / R \gtrsim 0.2$. Since transverse expansion leads to faster dilution compared to the $0+1$-D case, the system is driven again towards free streaming and $dE_{\rm tr} / d\eta$ stops decreasing. On the other hand, as the initial fireball falls apart, its ellipticity quickly decays as it is converted to elliptic flow.

\section{Final-state observables}

\begin{figure}
\begin{center}
\begin{tabular}{cc}
 \includegraphics[width=0.48\linewidth]{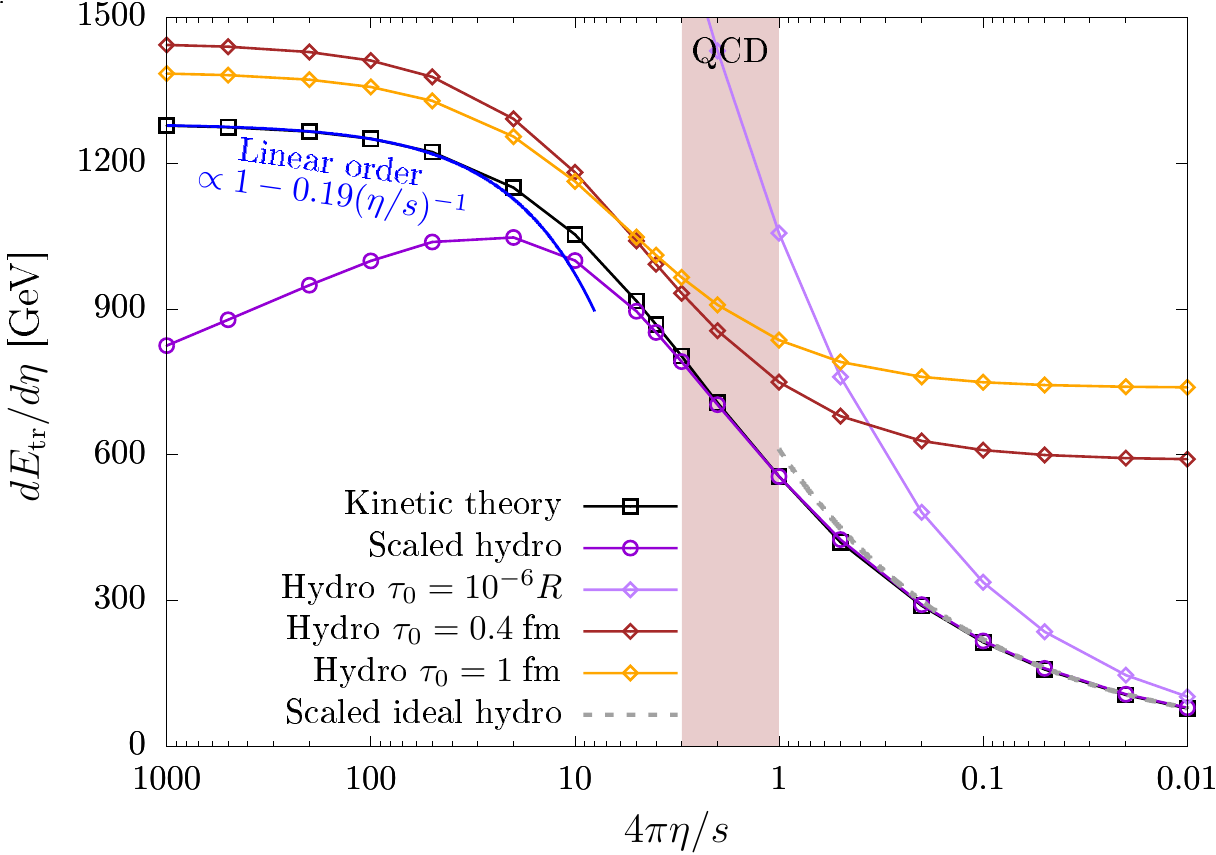} &
 \includegraphics[width=0.48\linewidth]{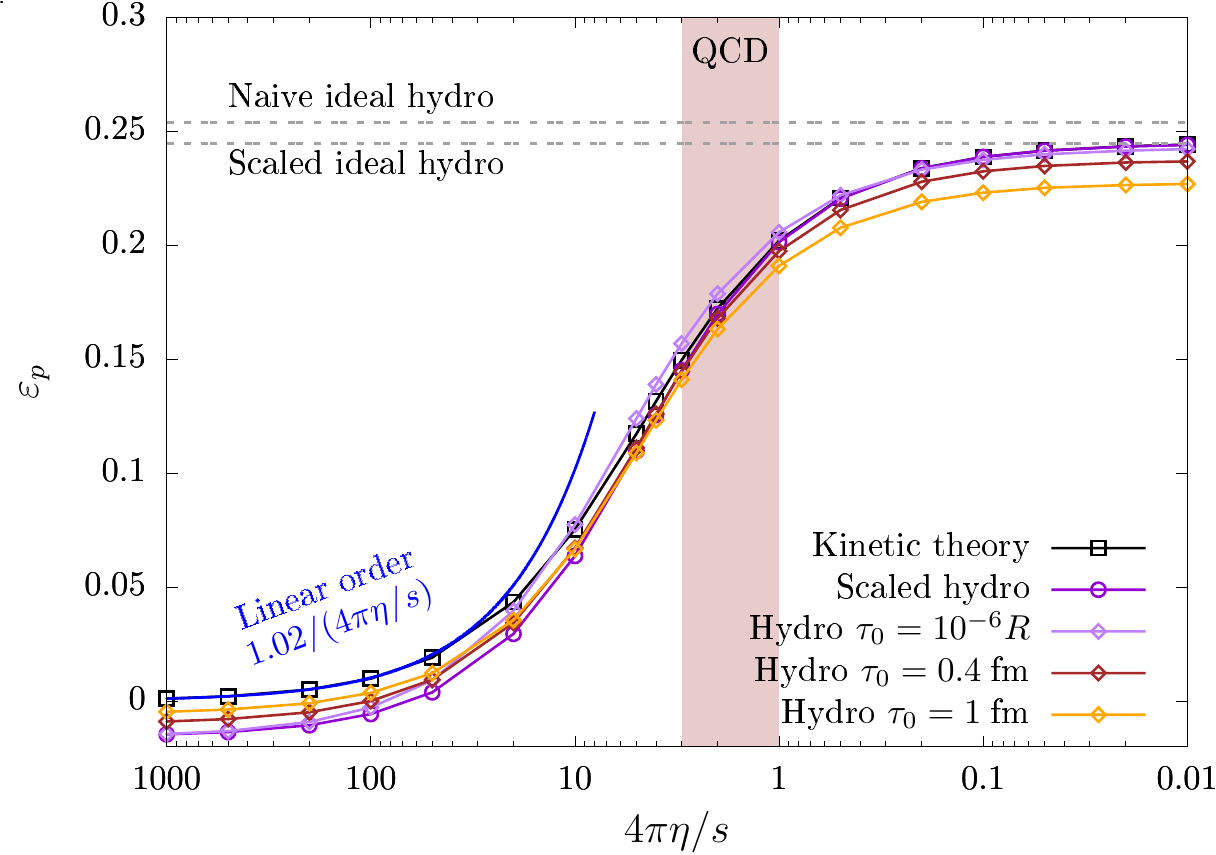}
\end{tabular}
\caption{
Transverse plane energy $dE_{\rm tr} / d\eta$ (left) and elliptic flow coefficient $\varepsilon_p$ (right), measured at $\tau = 4R$, represented with respect to $4\pi \eta / s$. Kinetic theory results (black lines) are compared with hydrodynamics results with scaled (purple) and naive initialization at $\tau_0 = 10^{-6} R$ (light purple), $0.4$ fm (brown) and $1$ fm (orange). The gray dotted line in the left panel shows the expected ideal hydrodynamics scaling, $dE_{\rm tr} / d\eta \sim (4\pi\eta / s)^{4/9}$.\vspace{-10pt}
\label{fig:transverse}
}
\end{center}
\end{figure}

We now look at the final-state observables, measured at $\tau = 4R$ for various values of $4\pi \eta / s$. The left panel of Fig.~\ref{fig:transverse} shows $dE_{\rm tr} / d\eta$, which decreases in RTA (black line) as $(4\pi \eta / s)^{4/9}$ close to the perfect fluid limit (indicated by the gray dashed line). This trend is captured very well by hydrodynamics with scaled initialization. Conversely, the naive initalization (shown with the purple line) gives a $(4\pi \eta / s)^\gamma$ dependence, with $\gamma - 4/9 \simeq 0.08$, thus diverging from the RTA prediction as $4\pi \eta / s$ is increased. This diverging behaviour is partially cured by initializing at later times, thus omitting more and more of the unphysical early-time behavior. Looking now at the large $4\pi \eta / s$ limit, we see that the hydrodynamical results corresponding to the scaled initialization diverge from the RTA ones. This is because as $4\pi \eta / s$ increases, $\wt$ decreases throughout the system and the time required for equilibration ($\wt \simeq 1$) eventually exceeds the transverse expansion timescale, $\sim 0.2 R$. Since transverse expansion interrupts equilibration, hydrodynamics and RTA never come into agreement, as reflected in the final-state value of $dE_{\rm tr} / d\eta$. 

The right panel of Fig.~\ref{fig:transverse} shows the elliptic flow coefficient, $\varepsilon_p$. At small $4\pi \eta / s$, RTA and hydrodynamics should come into agreement. However, this agreement is only reached when considering the scaled initialization. The curves corresponding to the naive initialization fall below the RTA curve, since during preequilibrium, inhomogeneous cooling leads to a larger decrease of ellipticity $\epsilon_2$, which is later converted into a smaller elliptic flow signal. At large $4\pi \eta /s$, $\varepsilon_p$ tends to $0$ for RTA, while in hydrodynamics with both the scaled and the naive initializations, $\varepsilon_p$ becomes negative. 

\section{Conclusion}

The focus of this paper was on the effect of preequilibrium dynamics on final-state observables in Bjorken flow with transverse expansion. By studying the evolution along the attractor of kinetic theory (RTA) and hydrodynamics, we were able to pinpoint the source of discrepancies between the two theories. Specifically, we showed that in hydrodynamics, the transverse plane energy per rapidity, $dE_{\rm tr} / d\eta$, increases with time during preequilibrium, while for RTA, it stays constant. This leads to a significant overestimation of the final-state energy. Moreover, due to the inhomogeneous cooling of hotter and colder regions of the transverse plane, the eccentricity coefficients $\epsilon_n$ of the initial profile is modified during preequilibrium. In the case considered here, the hydrodynamical evolution leads to a larger decrease of $\epsilon_n$ compared to RTA. This in turn causes the hydrodynamical prediction for the final-state elliptic flow coefficient $\varepsilon_p$ to fall below that of RTA. 

To cure the above shortcomings of hydrodynamics, we rescaled the initial conditions for hydrodynamics, such that hydrodynamics and RTA agree during the late-time, equilibrated phase of Bjorken flow. This rescaling brings the hydrodynamics and RTA results into agreement for $4\pi \eta / s \lesssim 3$, which includes the regime relevant for strongly-interacting fluids (QCD). At larger values of $4\pi \eta / s$, our rescaling scheme is no longer suitable, since equilibration is interrupted by transverse expansion and the fluid stays out of equilibrium throughout its entire evolution (see Ref.~\cite{Ambrus:2022} for an extended discussion).

\vspace{-10pt}
\begin{acknowledgments}
This work is supported by the Deutsche Forschungsgemeinschaft (DFG, German Research Foundation)
through the CRC-TR 211 ’Strong-interaction matter under extreme conditions’– project
number 315477589 – TRR 211. V.E.A.~gratefully acknowledges the support through a grant of the 
Ministry of Research, Innovation and Digitization, CNCS - UEFISCDI,
project number PN-III-P1-1.1-TE-2021-1707, within PNCDI III. 
\end{acknowledgments}

\vspace{-10pt}
\nocite{*}

\end{document}